\documentclass{elsart}

\usepackage{epsfig}

\begin{document}
\begin{frontmatter}
\title{Phase synchronization in tilted deterministic ratchets}

\author{Fernando R. Alatriste, Jos\'e L. Mateos\thanksref{EMAIL}}
\address{Instituto de F\'{\i}sica, 
Universidad Nacional Aut\'onoma de M\'exico, \\
Apartado Postal 20-364, 01000 M\'exico, D.F., M\'exico}
\thanks[EMAIL]{E-mail: mateos@fisica.unam.mx; \\
Fax: (52) (55) 5622 5015; Phone: (52) (55) 5622 5130}

\begin{abstract}

We study phase synchronization for a ratchet system. We consider
the deterministic dynamics of a particle in a tilted ratchet potential 
with an external periodic forcing, in the overdamped case. 
The ratchet potential has to be tilted in order to obtain a rotator or 
self-sustained nonlinear oscillator in the absence of external 
periodic forcing. This oscillator has an intrinsic 
frequency that can be entrained with the frequency of the external driving. 
We introduced a linear phase through a set of discrete time events
and the associated average frequency, and show that this frequency can be  
synchronized with the frequency of the external driving.
In this way, we can properly characterize the phenomenon
of synchronization through Arnold tongues, which represent regions of 
synchronization in parameter space, and discuss their implications 
for transport in ratchets. 
                           
\end{abstract}

\end{frontmatter}

\section{Introduction}

The phenomenon of synchronization is widespread in Nature.
We witness its manifestations in many different places and 
contexts. Synchronization is essentially a nonlinear phenomenon
and is very common in many complex systems, not only in the physical
sciences, but in the life sciences as well \cite{piko,glass,bocca,acebron}. 
In particular, the case of phase synchronization establishes a common
formalism to treat both nonlinear periodic oscillators, as well as chaotic 
and noisy oscillators \cite{piko,ani}.

In a different context, there has been an increasing interest during recent
years in the study of transport phenomena of nonlinear
systems that can extract usable work from unbiased non-equilibrium
fluctuations. These, so called Brownian motors 
(or thermal ratchets) can be modeled by a Brownian particle undergoing
a random walk in a periodic asymmetric potential, and being acted upon
by an external time-dependent force of zero average. The recent burst
of work is motivated by both, (i) the challenge to model
unidirectional transport of molecular motors within the biological
realm and, (ii) the potential for novel technological applications
that enables an efficient scheme to shuttle, separate and pump 
particles on the micro- and even nanometer scale 
\cite{rev1,rev2,rev3,rev4,rev5,rev6}.

Although the vast majority of the literature in this field considers
the presence of noise, there have been attempts  to model the
transport properties of classical deterministic ratchets as well
\cite{bart,ajdari,han96,jun,ma1,ma2,ma3,fabio02}. In this paper we will 
be dealing with a deterministic tilted ratchet in the overdamped regime that 
acts as a rotator or self-sustained oscillator with a characteristic
frequency, even in the absence of an external periodic forcing.
The dynamics can be represented by a particle in a washboard
potential that has been studied in many different contexts, like phase
dynamics in synchronization \cite{piko,ani}, pendulum dynamics
\cite{baker}, rotators \cite{piko}, superionic conductors \cite{risken},
optical potentials \cite{kishan}, excitable systems \cite{lin}, 
diffusion on surfaces \cite{katja1,katja2,guantes1,guantes2,reich04,save05},
charge density waves \cite{bohr} and Josephson junctions dynamics 
\cite{piko,risken,fabio05}. When the washboard potential is periodically 
driven it exhibits a great variety of nonlinear phenomena including phase 
locking, hysteresis \cite{fabio99} and chaos \cite{kautz}.   
 
Here we will study the synchronization properties of an overdamped
particle moving on a tilted ratchet potential that is rocked by a
periodic external force. Throughout this paper we will consider a
constant force above the critical value, in such a way that the particle
slides down the washboard potential, even though it is in the 
overdamped regime. When the periodic forcing is absent, 
the particle experiences only a fixed washboard potential,
and moves through each period of the ratchet in a given constant 
time that defines the period $\tau_0$ of this rotator.
The asociated frequency $\omega_0 = 2\pi /\tau_0$ is its 
characteristic frequency. 
In this sense, this rotator is effectively acting as a self-sustained oscillator,
with its own characteristic frequency. That is, if the rotator is driven by a
constant force, it acquires the same features of a self-sustained oscillator,
having a limit cycle in phase space. Thus, forced rotators are similar to 
self-sustained oscillators and can be synchronized by a periodic external force.   
We will drive this rotator with an external periodic force of period
$\omega_D$. In this way we can define properly the synchronization 
of the rotator and the external forcing. The current or average velocity,
which is the important quantifier for this system, displays 
steps as a function of a control parameter. This last
result has been found previously by other authors that have studied
overdamped deterministic ratchets \cite{bart,ajdari,han96,rubi}. 
On the experimental side, these so called Shapiro steps have been 
found recently for deterministic Josephson vortex ratchets and
three-junction SQUID rocking ratchets \cite{koelle1,koelle2}.

More recent studies consider the problem of synchronization of 
deterministic ratchets, but they deal with complete synchronization 
between two coupled ratchets \cite{vinc1,vinc2,vinc3,family}, 
and with anticipated synchronization between two unidirectional 
coupled ratchets with time delay \cite{kostur}. 
In this work, instead, we are dealing with phase synchronization
through a linear phase, properly defined through a set of discrete
time events. 
        
As a way of characterizing the synchronization phenomenon, 
we will calculate, for the first time, the so called Arnold tongues for 
the tilted deterministic ratchet. For a description of Arnold tongues
in circle maps and pendulum dynamics see \cite{baker,bohr,jensen}. 
Arnold tongues are regions of synchronization in a parameter space.
Here we will calculate these regions in a two-dimensional parameter 
space defined by the ratio $\omega_D /\omega_0$ and the amplitude
of the driving periodic force $F_D$. The tips of the tongues are
located on rational values of the ratio $\omega_D /\omega_0 = p/q$, 
where $p$ and $q$ are integer numbers. Each tongue is therefore 
labeled by a rational $p/q$ whose inverse is precisely the value of the current 
in the driven washboard potential; the widths of these Arnold tongues 
correspond to the size of the steps of the current as a function of $F_D$.

\section{Tilted ratchets as nonlinear rotators}

To start out, let us consider now the one-dimensional problem of a
particle driven by a periodic time-dependent external force in an
asymmetric periodic ratchet potential. Here, we do not
take into account any sort of noise, meaning that the dynamics is
deterministic. Two additional forces act on the particle: a dissipative force 
proportional to velocity, and an external constant force. 
We thus deal with a rocked deterministic tilted ratchet\cite{bart,han96} 
in the overdamped limit that obeys the following equation of motion:

\begin{equation}
m\gamma \dot x + {\frac{dV(x)}{dx}} = F + F_D \cos(\omega_D t),
\end{equation}

\noindent where $m$ is the mass of the particle, $\gamma$ is the friction coefficient,
$V(x)$ is the asymmetric periodic ratchet potential, $F$ is a constant force, 
$F_D$ and $\omega_D$ represent the amplitude and the frequency of the external
driving force, respectively. The ratchet potential is given by 

\begin{equation}
V(x) = V_0 \left [C - \sin {\frac{2\pi (x-x_0)}{L}} - {\frac{1}{4}}\sin {\frac{4\pi (x-x_0)}{L}} \right ],
\end{equation}

\noindent where $L$ is the periodicity of the potential, 
$V_0 $ is the amplitude, and $C$ is an arbitrary constant. The potential
is shifted by an amount $x_0$ in order that the minimum of the potential
is located at the origin \cite{ma1}.

Let us define the following dimensionless units: $x^{\prime }=x/L$,
$x_{0}^{\prime }=x_{0}/L$, $t^{\prime } = \gamma t$,
$\omega_{D}^{\prime }=\omega _{D}/\gamma$,  
$F^{\prime }=F/mL\gamma^{2}$,  
$F_{D}^{\prime }=F_{D}/mL\gamma^{2}$, $V^{\prime } = V /mL^{2}\gamma^{2}$
and $V_{0}^{\prime } = V_{0} /mL^{2}\gamma^{2}$.
Thus, we are using the periodicity of the potential $L$ as the natural length scale and 
the inverse of the friction coefficient $\gamma$ defines the natural time scale. With these
two quantities, the natural force is given by $mL\gamma^{2}$ and the associated energy
by  $mL^{2}\gamma^{2}$.

The dimensionless equation of motion, after renaming the variables again without the 
primes, becomes 

\begin{equation}
\label{emov}
\dot{x}+{\frac{dV(x)}{dx}}= F + F_D \cos (\omega_D t),   
\end{equation}

\noindent where the dimensionless potential can be written as

\begin{equation}
V(x) = V_0 \left [C - \sin 2\pi (x-x_{0}) - {\frac{1}{4}} \sin 4\pi (x-x_{0}) \right ]
\end{equation}

\noindent and is depicted in the inset of Fig. 1. The constant $C$ is
such that $V(0)=0$, and is given by 
$C = -(\sin 2\pi x_{0} + 0.25 \sin 4\pi x_{0})$.
We choose, $x_{0}\simeq -0.19$

We can rewrite the equation of motion Eq. (3) as

\begin{equation}
\dot{x} + {\frac{\partial U(x,t)}{\partial x}} = 0,
\end{equation}

\noindent where $U(x,t) = V(x) - [F + F_D \cos(\omega_D t)]x$.
The important point to stress here is that we need to add a constant force $F$ to the 
ratchet in order to tilt the ratchet potential and, in this way, obtain a rotator 
(that acts as a self-sustained oscillator),
even without the external periodic forcing \cite{piko,osipov}. In this way, we can 
properly synchronize the characteristic frequency of the rotator with the driving frequency 
$\omega_D$. 

\begin{figure}
\begin{center}
\epsfig{figure=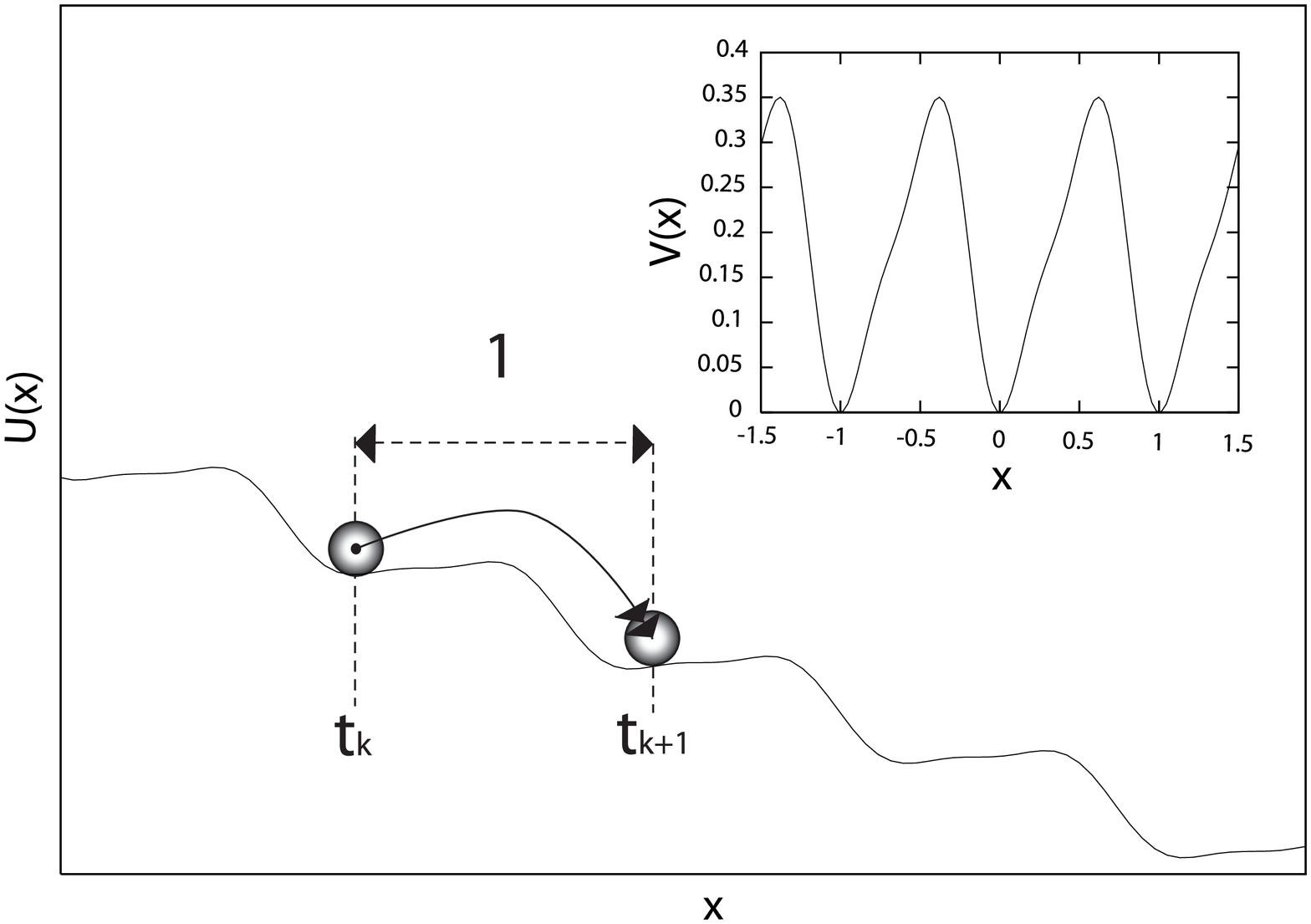,width=12.0cm}
\end{center}
\caption{The tilted washboard potential indicating the dynamics that defines the
discrete time events $t_k$. The ratchet potential without tilt is illustrated in the inset.}
\label{fig1}
\end{figure}

When $F_D = 0$, we have a tilted ratchet that obeys the equation of motion:
$\dot{x} + dV(x)/dx = F$. The tilted (time-independent) washboard potential is, 
in this case, $U(x) = V(x) - Fx$, see Fig. 1.
Thus, this ratchet becomes a rotator that has a characteristic
frequency $\omega_0$. The associated period of the rotator $\tau_0 = 2\pi /\omega_0$ 
can be obtained directly by integrating this equation of motion.
However, there is another way to obtain $\omega_0$ for this tilted ratchet
that relays on the introduction of a phase variable for this rotator.
So, in what follows we will introduced this general concept that we will use in 
the rest of the paper.

In order to define a phase variable we need first to obtain a discrete process from the 
continuous dynamics by introducing discrete time events. These discrete times can be defined
as the times when the particle arrives at the discrete position $x_k = \pm k$, which correspond
to the minima of the ratchet potential without tilt. Here $k = 0,1,2,...$. Remember that the
period of the ratchet is one: $V(x+1) = V(x)$.
This defines the set of times ${t_k}$, where $k$ is a nonnegative integer. In Fig. 1 we show the
washboard potential illustrating these discrete times. Once we obtain these set of markers, we
can define an instantaneous linear phase for the rotator as \cite{piko,ani}

\begin{equation} 
\phi (t) = 2\pi {\frac{t-t_{k}}{t_{k+1}-t_{k}}} + 2\pi k, \qquad t_{k} \le t < t_{k+1}
\end{equation}

This linear phase is valid in the indicated interval and defines a piecewise linear function
of time that increases by $2\pi$ each time the particle crosses the dimensionless 
position $x_{k} = \pm k$.

Given this phase, we can define the instantaneous frequency of the rotator 
as $\omega (t) = \dot{\phi} (t)$, and the average frequency as

\begin{equation} 
\langle \omega \rangle = \lim_{T\to \infty} {\frac{1}{T}} \int_{0}^{T} \omega (t) dt
= \lim_{T\to \infty} {\frac{1}{T}} [\phi(T) - \phi (0)]. 
\end{equation}

Without loss of generality, we choose $t_{0} = 0$, and thus $\phi (0) = 0$. 
The limit above can be written as 

\begin{equation}
\label{fre} 
\langle \omega \rangle = \lim_{k\to \infty} {\frac{\phi(t_k)}{t_k}} 
= 2\pi \lim_{k\to \infty} {\frac{k}{t_k}}
\end{equation}

This is the simplest way to calculate the average frequency; we simply count the number of
jumps (given by $k$) and divide by the total time span $t_k$.

The other quantity of importance is the average velocity or current in the tilted ratchet.
In order to evaluate this current we have to calculate the number $k$ of unit periods 
that the particle crosses to the right, denoted by $N^{R}_k$, and the number of crossings
to the left, given by $N^{L}_k$. The total number of periods traversed on the ratchet is given by 
$N^{T}_k = N^{R}_{k} + N^{L}_{k}$. The difference  

\begin{equation}
N_k = N^{R}_{k} - N^{L}_{k}
\end{equation}

indicates that during the time $t_k$ the particle has covered the distance $x_k = N_k$.

Therefore, the average velocity (current) is given by

\begin{equation} 
\label{cur}
\langle v \rangle = \lim_{k\to \infty} {\frac{N_k}{t_k}} 
\end{equation}

In the particular case when $N^{L}_k = 0$, that is, when there are no jumps to the left,  
we have $N_k = N^{R}_{k} = k$. Thus,

\begin{equation} 
\langle v \rangle = \lim_{k\to \infty} {\frac{k}{t_k}} = {\frac{1}{2\pi}} \langle \omega \rangle
\end{equation}

In the simple case of a tilted ratchet without external forcing, $F_D = 0$, 
the average frequency defined above coincides with the natural frequency of the rotator, that is,
$\langle \omega \rangle = \omega_0$.  

The above treatment is quite general and can be used in the case of a tilted ratchet 
with an inertial term, even though this inertial ratchet can display a chaotic dynamics 
\cite{jun,ma1,ma2,ma3,fabio02}, and also in the case of a tilted ratchet with noise. 
We have used the concept of a linear phase due to its broad-range applications to
the cases of periodic and chaotic oscillators \cite{piko}, chaotic rotators \cite{osipov},
and oscillators in the presence of noise \cite{ani}. Additionally, the introduction of the
discrete events that define the linear phase allow us to simplify the dynamics and have
a more clear picture of the synchronization involved. 

\section{Numerical results}

\begin{figure}
\begin{center}
\epsfig{figure=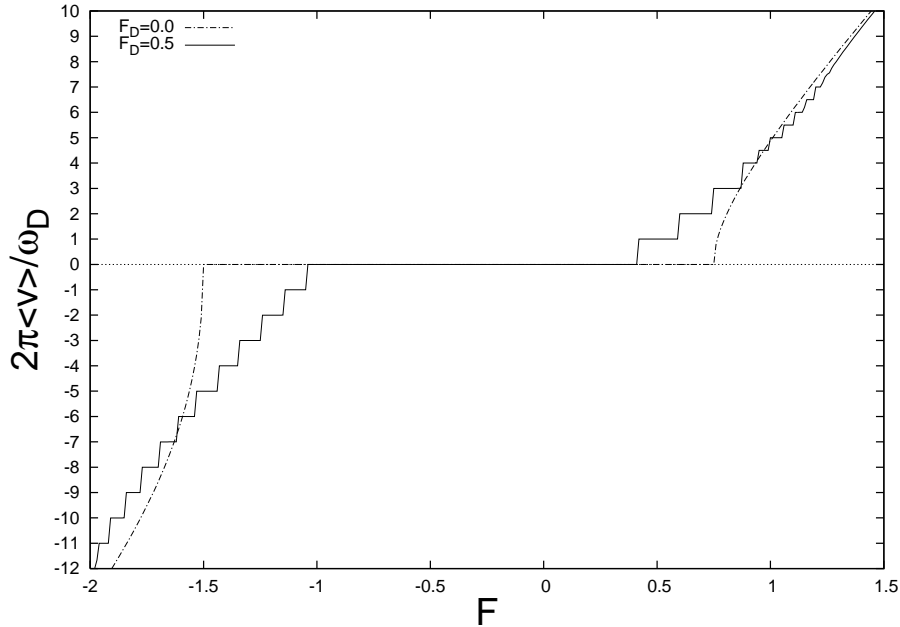,width=12.0cm}
\end{center}
\caption{The average velocity as a function of the external force $F$. 
The dashed line indicates the case when the periodic driving is absent 
$(F_{D} = 0$) and the continuous line is the case when the periodic 
driving is acting on the particle with an amplitude $F_{D} = 0.5$. 
In both cases we used $\omega_{D} = 0.7$.}
\label{fig2}
\end{figure}

In this section we will solve numerically the equation of motion for the rocking tilted ratchet.
We use the fourth-order Runge-Kutta algorithm to solve the differential equation Eq. (\ref{emov}). 
Once we obtain the full trajectory, we identify the set of discrete times ${t_k}$ when the
particle crosses the positions $x_k$. With this marker events we calculate directly the
average frequency using Eq. (\ref{fre}) and, after calculating the quantity $N_k$, we obtain 
the current, using Eq. (\ref{cur}). We will fix throughout the paper the amplitude of the ratchet 
potential as $V_0 = 1/2\pi$. With this value, the critical tilt to the right is 
$F^R_{c} = 0.75$ and to the left $F^L_{c} = -1.5$. 

\begin{figure}
\begin{center}
\epsfig{figure=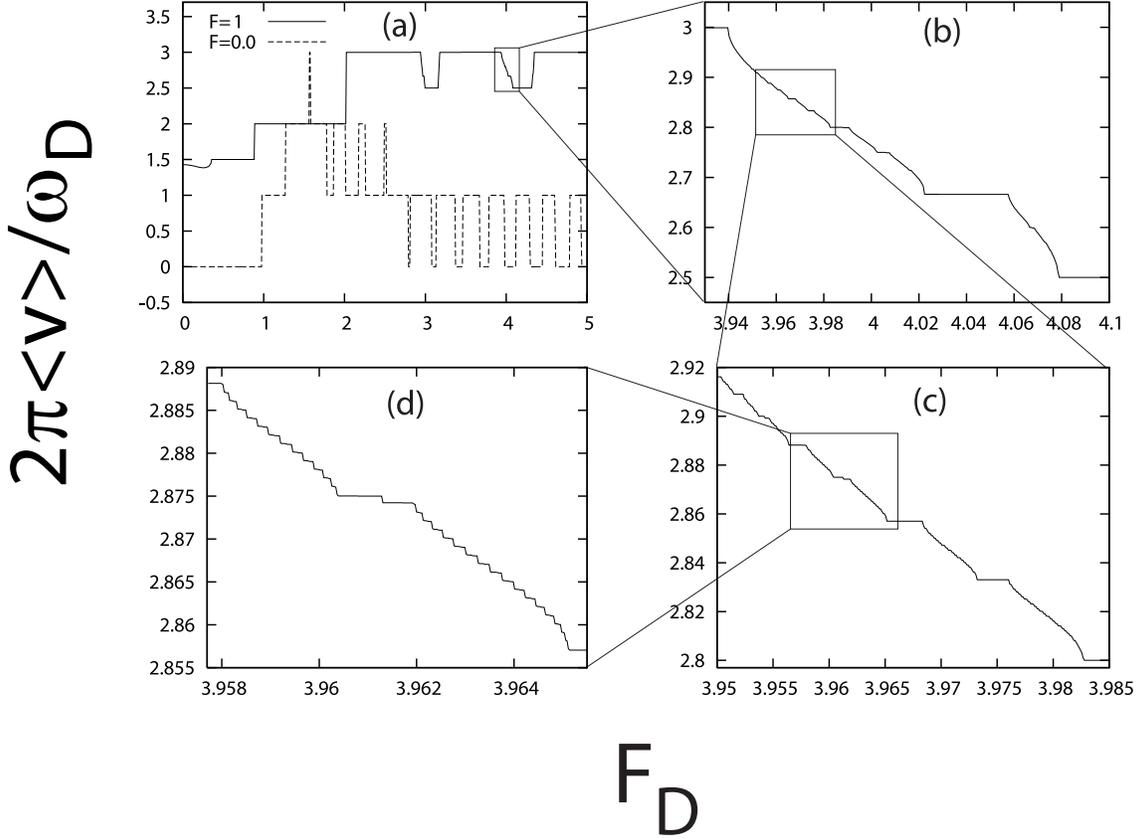,width=15.0cm}
\end{center}
\caption{The scaled average velocity $2\pi \langle v \rangle /\omega_D$
as a function of the amplitude of the periodic driving $F_D$. 
(a) The dashed line indicates the case when the tilt is absent
$(F = 0$) and the continuous line is the case when the tilt is $F = 1$.
(b), (c) and (d) correspond to successive magnifications of the current
for the tilted ratchet, showing a self-similar structure of steps, typical
of a devil's staircase. In the dashed line in (a), we used 
$\omega_{D} = 0.7$, since in this case $\omega_0$ is not defined.
In the other cases we used $\omega_{D} = 0.7 \omega_0$.}
\label{fig3}
\end{figure}

In Fig. 2, we depict the average velocity, scaled with the driving frequency, 
$2\pi \langle v \rangle /\omega_D$ as a function of the tilt $F$. The dashed line
shows the case without periodic driving ($F_{D} = 0$) and corresponds to the
fixed washboard potential $U(x) = V(x) - Fx$. Notice that the current is zero until
we arrive at the critical tilt $F^R_{c} = 0.75$ to the right or to the critical tilt 
$F^L_{c} = -1.5$ to the left. This step of zero current is not centered around the origin,
due to the asymmetry of the ratchet potential. For values greater than $F^R_{c}$ we have a finite 
current that increases monotonically with $F$. Of course, for values less than
$F^L_{c}$ we obtain a negative average velocity that decreases for negative 
values of the tilt. When the periodic driving is present, this scaled current acquires a
series of clearly defined steps for values of the current given by the ratio $p/q$,
where $p$ and $q$ are integer numbers. In many cases, $q = 1$ and the 
average scaled current is an integer. In the context of Josephson junctions, 
these are the celebrated Shapiro steps \cite{kautz}.

Remember that in the case where all the jumps are to the right, that is, 
$N^{L}_k = 0$, we show that $2\pi \langle v \rangle = \langle \omega \rangle$.
Therefore, a rational value of $2\pi \langle v \rangle /\omega_D = p/q$ means that
$\langle \omega \rangle = (p/q) \omega_D$ for a whole range of values of the tilt. 
This phenomenon is called frequency locking.

In Fig. 3, we show the scaled average velocity $2\pi \langle v \rangle /\omega_D$
as a function of the amplitude of the periodic driving $F_D$. In (a), the
dashed line depicts the current for the ratchet without tilt $(F = 0)$ and coincide
with previous calculations \cite{bart,ajdari,han96} showing a structure of steps
of unit height. The solid line shows the current for a tilted ratchet with $F = 1$
that also has well defined steps for rational values. In (b), (c) and (d) 
we show successive magnifications of the current that clearly exhibits a 
self-similar structure of steps, typical of a devil's staircase \cite{jensen,reich99}. 
This detailed structure has been reported before \cite{ajdari,han96} for an 
overdamped ratchet without tilt, but here we obtained this fractal current 
with a devil's staircase also for the tilted ratchet.     

\begin{figure}
\begin{center}
\epsfig{figure=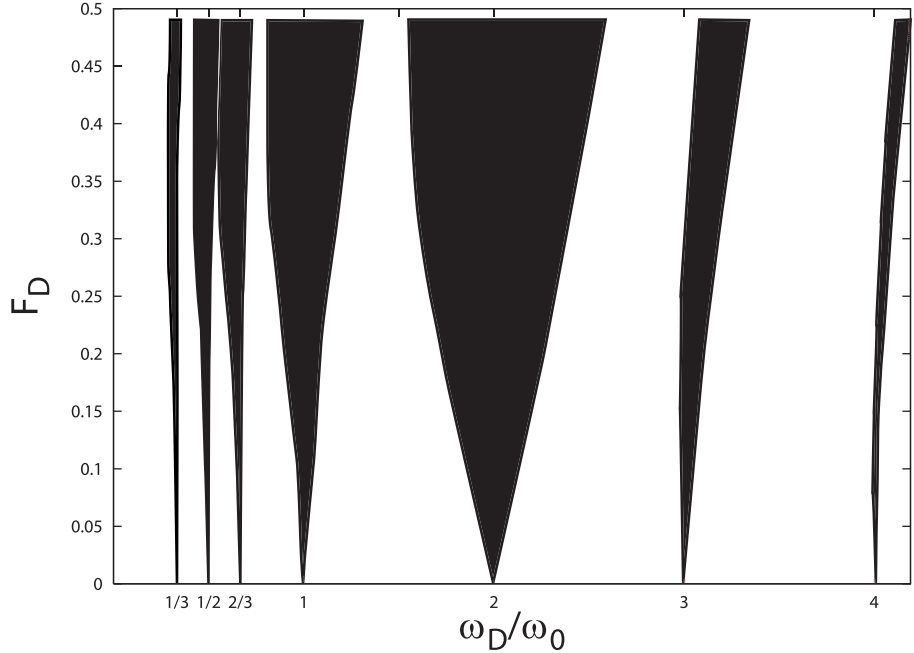,width=12.0cm}
\end{center}
\caption{Arnold tongues in the parameter space $F_D$ against $\omega_D/\omega_0$.
Notice that the tongues are located precisely at the rational values $p/q$, for $p$ and $q$ integer
numbers, of the ratio between the driving frequency and the natural frequency of the rotor.
Here the tilt is $F = 1$ and the corresponding frequency is $\omega_0 \simeq 3.41$ . 
Each tongue is labeled by the inverse $q/p$ that gives the value of the scaled
average velocity $2\pi \langle v \rangle /\omega_D$ in that region of the parameter space.}
\label{fig4}
\end{figure}

In Fig. 4, we depict the parameter space $F_D$ against $\omega_D/\omega_0$
that shows regions of synchronization, called Arnold tongues, located at 
rational values $p/q$, where $p$ and $q$ are integer numbers. 
We choose a tilt $F = 1$, which corresponds to $\omega_0 \simeq 3.41$.
Notice that the tongues start, for small values of $F_D$, precisely at these rational values,
as indicated in the figure. Each Arnold tongue corresponds to one 
particular rational $p/q$ whose inverse gives the value of the scaled average velocity
$2\pi \langle v \rangle /\omega_D$ in that region of the parameter space.
This is depicted in a three-dimensional plot in Fig. 5. Therefore, the average
velocity, properly scaled, acquires rational values $q/p$, which correspond to
the phenomenon of phase synchronization in this forced tilted ratchet, 
acting as a rotator \cite{osipov}. 

\begin{figure}
\begin{center}
\epsfig{figure=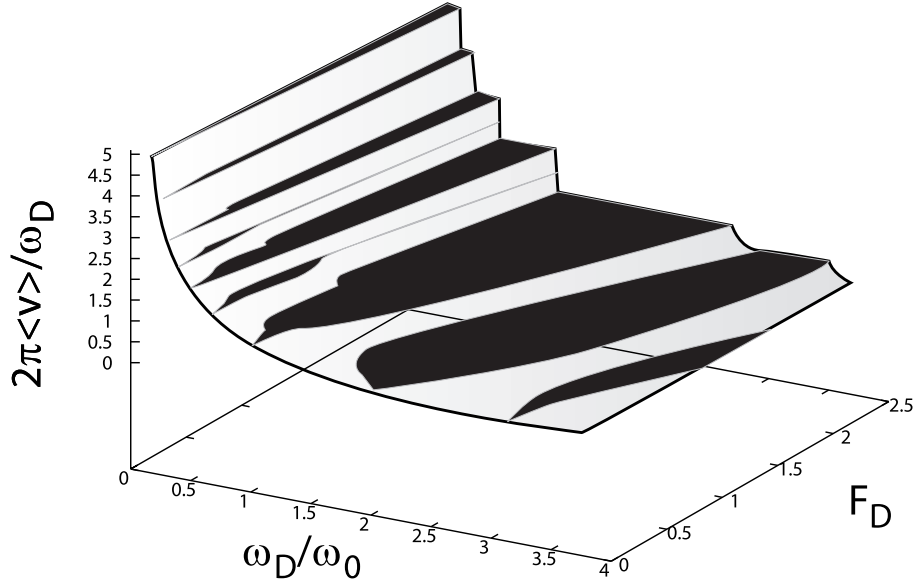,width=12.0cm}
\end{center}
\caption{Three-dimensional plot of the scaled average velocity 
$2\pi \langle v \rangle /\omega_D$ as a function of $F_D$ and $\omega_D/\omega_0$.
A projection of this 3D plot shows the Arnold tongues in the parameter space of Fig. 4.
Here the tilt is $F = 1$ and $\omega_0 \simeq 3.41$.}
\label{fig5}
\end{figure}

\begin{figure}
\begin{center}
\epsfig{figure=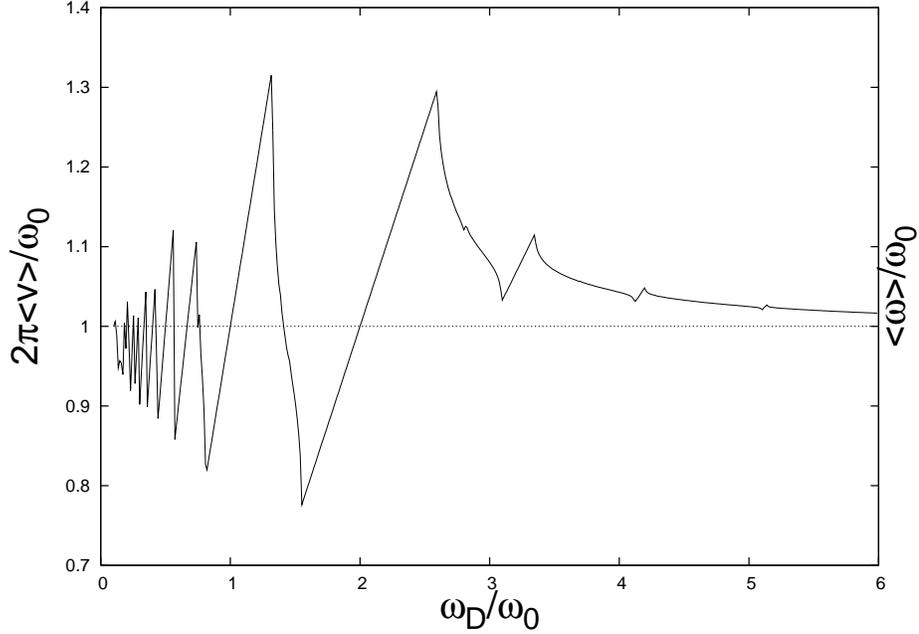,width=12.0cm}
\end{center}
\caption{Average velocity, scaled now as $2\pi \langle v \rangle /\omega_0$, 
and the average frequency of the rotator, scaled as $\langle \omega \rangle /\omega_0$
as a function of the ratio $\omega_D/\omega_0$. The straight line segments 
correspond to the steps observed due to phase synchronization in Fig. 4.
The peaks in the current are located at the right borders of the Arnold tongues
in Fig. 4. Here the tilt is $F = 1$, $\omega_0 \simeq 3.41$, and $F_D = 0.5$.}
\label{fig6}
\end{figure}

Finally, in Fig. 6, we plot the average velocity, scaled now as 
$2\pi \langle v \rangle /\omega_0$, as a function of the ratio $\omega_D/\omega_0$. 
Here $\omega_0$ is a fixed value that correspond to the characteristic 
frequency of the tilted ratchet in the absence of periodic driving.
In the same figure, we plot the average frequency of the rotator, scaled as
$\langle \omega \rangle /\omega_0$, calculated using the discrete dynamics 
explained in the previous section.
Instead of plateaus, we have now straight lines, since we are scaling the 
current with a fixed parameter $\omega_0$, instead of the running parameter 
$\omega_D$. In this case, all the jumps are to the right direction and 
therefore $2\pi \langle v \rangle = \langle \omega \rangle$, that is, the average velocity
is proportional to the average frequency of the rotator. 

Notice that the peak values of the average velocity correspond to the borders of the 
Arnold tongues in Fig. 4. Thus, when we are crossing an Arnold tongue 
(synchronization region) the current increases linearly with $\omega_D/\omega_0$;
at the right border of the tongue the current is maximal and outside the tongue 
starts to decrease until we arrive at the next tongue to increase linearly again, and so
forth. At the right border of the Arnold tongue, labeled by $p/q = 1$, the current has
a maximum, followed by the second largest peak at the right border of the tongue
with $p/q = 2$. The heights of the peaks in the current arise due to the combined 
effect of both the width of the tongues and the slope of the linear segments. The slopes
of each of the segments correspond precisely to the inverse values $q/p$ that label the tongues. 
Therefore, the peaks in the current are associated with the phenomenon of synchronization.

\section{Concluding remarks}

In summary, we have analyzed the phenomenon of phase synchronization in tilted
deterministic ratchets in the overdamped regime and with an external periodic forcing.
The dynamics in this rocked washboard potential corresponds precisely with the
dynamics of a rotator that, acting as a self-sustained oscillator, can be capable of 
being synchronized with the external periodic drive. We can clearly identify three
frequencies for this system: the characteristic frequency of the rotator without driving,
the driving frequency itself, and the average frequency of the rotator with driving. 
This average frequency is the derivative of a time-dependent phase,
that can be obtained through a set of discrete time events and is a piecewise 
linear function of time between these markers. We calculated the average frequency
and the average velocity as a function of the tilt and obtained the
well-know Shapiro steps that characterize the phenomenon of frequency locking.
We also exhibit the self-similar structure of steps in the current, typical of a
devil's staircase. We obtained well-defined Arnold tongues in the 2D parameter 
space given by the amplitude and the frequency of the periodic forcing. Each Arnold tongue 
is labeled by a rational number $p/q$, where $p$ and $q$ are integer numbers, whose inverse gives 
precisely the rational value of the average scaled velocity of the particle. Finally, we
show that the local maxima in the average velocity correspond to the borders of these
Arnold tongues and, in this way, we established a connection between optimal
transport in ratchets and the phenomenon of phase synchronization. 
  
\bigskip
{\bf Acknowledgements}
\medskip

FRA gratefully acknowledges financial support from CONACYT scholarship. 
JLM also wants to thank the Alexander von Humboldt Foundation for support.

\end{document}